\documentclass[journal]{IEEEtran}

\usepackage{xcolor,soul,framed} %,caption

\colorlet{shadecolor}{yellow}
\usepackage[pdftex]{graphicx}
\graphicspath{{../pdf/}{../jpeg/}}
\DeclareGraphicsExtensions{.pdf,.jpeg,.png}

\usepackage[cmex10]{amsmath}
\usepackage{array}
\usepackage{mdwmath}
\usepackage{mdwtab}
\usepackage{eqparbox}
\usepackage{url}
\usepackage{stfloats}
\usepackage{algorithm}
\usepackage{algorithmic}
\usepackage{adjustbox}
\usepackage{tabularx} % for 'tabularx' environment and 'X' column type
\usepackage{booktabs} % for \toprule, \midrule, and \bottomrule
\usepackage{amssymb}
\begin{document}
\bstctlcite{IEEEexample:BSTcontrol}
    \title{Generating HSR Bogie Vibration Signals via Pulse Voltage-Guided Conditional Diffusion Model}
    
    \author{Xuan~Liu,
    Jinglong~Chen,~\IEEEmembership{Member,~IEEE,} Jingsong~Xie, Yuanhong Chang\\}

% ====================================================================
\maketitle

% === ABSTRACT ====================================================================
% =================================================================================
\begin{abstract}
%\boldmath
Generative Adversarial Networks (GANs) for producing realistic signals, have substantially improved fault diagnosis algorithms in various Internet of Things (IoT) systems. However, challenges such as training instability and dynamical inaccuracy limit their utility in high-speed rail (HSR) bogie fault diagnosis. To address these challenges, we introduce the Pulse Voltage-Guided Conditional Diffusion Model (VGCDM). Unlike traditional implicit GANs, VGCDM adopts a sequential U-Net architecture, facilitating multi-phase denoising diffusion for generation, which bolsters training stability and mitigate convergence issues. VGCDM also incorporates control pulse voltage by cross-attention mechanism to ensure the alignment of vibration with voltage signals, enhancing the Conditional Diffusion Model's progressive controlablity. Consequently, solely straightforward sampling of control voltages, ensuring the efficient transformation from Gaussian Noise to vibration signals. This adaptability remains robust even in scenarios with time-varying speeds. To validate the effectiveness, we conducted two case studies using SQ dataset and high-simulation HSR bogie dataset. The results of our experiments unequivocally confirm that VGCDM outperforms other generative models, achieving the best RSME, PSNR, and FSCS, showing its superiority in conditional HSR bogie vibration signal generation. For access, our code is available at \underline{https://github.com/xuanliu2000/VGCDM}.
\end{abstract}

% === KEYWORDS 

% \fi
\IEEEpeerreviewmaketitle

% === I. INTRODUCTION ============================================================
\section{Introduction}

\IEEEPARstart The advent and widespread adoption of Internet of Things (IoT) technologies have transformed the landscape of modern railway systems, especially in the domain of high-speed rail (HSR) operations. These technologies facilitate the deployment of an array of sensors, which meticulously monitor and record numerous internal variables across subsystems. Parameters such as current and voltage, along with interactions with the external environment like knock and vibration, are tracked to provide real-time feedback on train performance. Vibration signals, emerging from the sophisticated interaction of machinery components, offer a holistic insight into both the internal mechanisms and external factors. This comprehensive perspective renders them invaluable for early malfunction detection\cite{ahmed2020condition}. However, maximizing the potential of these intricate vibration signals demands advanced analytical techniques. This ensures precise data alignment and seamless integration of diverse data types. The challenge here stems from the imperative to correlate varied operational parameters with specific mechanical behaviors. Traditional methodologies have predominantly leaned on parameters like rotational speed to discern malfunction patterns through time-frequency analysis. Nonetheless, their efficacy is circumscribed to certain conditions and they often fall short in tracking the evolving faulty features with the shift in operational scenarios.\cite{tian2015motor}\cite{Current-Aided}. 

The recent surge in research has underscored data-driven methodologies for vibration signal analysis, with a pronounced focus on machine learning and deep learning paradigms\cite{lei2020applications}\cite{9237134}. These approaches, including support vector machine(SVM)\cite{9464330}, convolutional neural network(CNN)\cite{8684317}, residual neural network(ResNet)\cite{10063190}, Transformer\cite{liu2023few} and etc. can extract latent features directly and automatically from measurements to assess the health-state. For instance, Jiang et al.\cite{Multimodal_jiang} use two deep belief networks(DBNs) separately extract the features of vibration and current signals, and pass through the fused feature into the third DBN for fault diagnosis. Wang \cite{8957109} combined a long short-term memory recurrent neural network and proactive model for the maintenance for HSR Power Equipment. These methods excel in handling complexity of diverse device data and achieve excellent performance in recognizing and detecting known patterns. However, these methods suffer from certain challenges of generalization under the real-world variable conditions found in HSR systems.  

To address these issues, several studies have explored the use of strategies improved by Variational Auto-Encoders (VAEs), and Generative Adversarial Networks (GANs)\cite{lu2017fault}\cite{pan2022generative}. By learning and mimicking the statistical properties of the training data, it is possible to generate new data that is similar to the real dataset, which can improve diagnostic performance by simulating device behavior under various normal and abnormal conditions. For instance, Zhao\cite{zhaoEnhancedDatadrivenFault2019a} et al. introduced a VAE to a fault diagnosis framework for data amplification via vibration signal generation, further enhancing fault diagnosis performance. Wan\cite{wan_qscgan_2021} et al. proposed quick self-attention convolutional generative adversarial network(QSCGAN), which utilizes self-attention and spectral normalization to enhance the stability and convergence rate of GAN in Liquid Rocket Engine bearings signals generation. Liu\cite{liu2022imbalanced} et al. improved GAN with a multi-scale residual network and hybrid loss function to balance rolling bearing imbalanced fault data distribution. Despite these methods have enhanced diagnostic systems' ability to manage challenges of imbalance and limited samples, they remain inefficient and inapplicable under many circumstances\cite{zhang2020small}. The training of GANs usually requires a careful balancing of the competition between the generator and the discriminator, leading to pattern crashes and unstable training. In addition, due to GANs is implicit, these methods still struggle in integrating multi-modal information, which constrains their applicability under more complex and various operating conditions like time-varying or multi-parameter control.   
Diffusion Models(DMs) have emerged as a powerful and promising new generative framework, which can achieve better generation quality and stability than GANs\cite{qu2023exploring}. Moreover, the success of DMs in text-to-image synthesis tasks, where the model generates high-fidelity images solely from textual descriptions, provides valuable insights for enhancing control in generative tasks\cite{xu2023versatile}. Contrastive Language–Image Pre-training(CLIP)\cite{radford2021learning} presented by OpenAI jointly learns visual and language representations by estimating the cosine similarity, allowing models to understand images paired with natural language descriptions, achieving incredible performance in alignment of images and texts. Dhariwal \cite{dhariwal2021diffusion} et al. illustrated the superior image synthesis capabilities of diffusion models over GANs. Subsequent works, Stable Diffusion\cite{Rombach_2022_CVPR} further illustrate the usage of conditional prompts guidance can achieve high-quality image generation and ControlNet\cite{zhang2023adding} underscored the enhanced controllability and efficacy of diffusion models in producing more accurate images under description like texts.

Drawing inspiration from these models, the generation of HSR vibration signals can also be steered by latent condition guidance, such as motor pulse voltage signals. Consequently, we explore the latent potential of leveraging pulse voltage signals to achieve direct and accurate vibration signal generation. We propose pulse \textbf{V}oltage \textbf{G}uided-\textbf{C}ondition \textbf{D}iffusion \textbf{M}odel (\textbf{VGCDM}), which encodes pulse voltage signals and align embedding by cross attention mechanism, enabling more accurate vibration signal generation in conditional diffusion model. The proposed model can effectively and controllable eliminate the imperceptible noise and take into account the periodic control information , which is suitable for the synthesis of vibration signals generation under variable conditions. Our work can further be a new and effective paradigm for enhancing monitoring and diagnosing HSR systems.

Furthermore, The main contributions of this work can be summarized in the following four points:

\begin{enumerate}
\item To our knowledge, our work pioneers the integration of control conditions (pulse voltage signals) for steering vibration signal generation of HSR system.
\item Leveraging the conditional diffusion models, our approach enhance the stability and controllablity in vibration signals generation.
\item We provide a comprehensive empirical evaluation, analyzing the performance of diffusion models in vibration signal generation and underscoring the method's merits, including validity under variable speed and real HSR scenario.
\end{enumerate}
 
In this paper, we explore using pulse voltage signals to guide the HSR bogie bearing vibration signals generation via conditional diffusion model. Particularly, related works are presented in Section II, and framework of the proposed method is introduced in Sections III. Furthermore, the proposed method is experimentally validated by the SQ and HSR bogie bearing dataset in section IV. Section V draws a discussion of proposed method. Ultimately, Section VI demonstrates the conclusion of entire approach.

% === II. Related works ========================
% =================================================================================
\section{Related works}

\subsection {Attention mechanism}
Attention mechanism\cite{bahdanau2014neural}, inspired by the human ability to selectively focus, have become an essential component of deep learning. Over the years, their inclusion has been instrumental in crafting many state-of-the-art models, spanning diverse applications\cite{vaswani2017attention}\cite{dosovitskiy2020image}\cite{liu2021swin}. Attention mechanism can be conceptualized as a method that associates a query with an array of key-value pairs to generate an output. By weighting the query against these pairs, it formulates a weight distribution across input components. It prioritizes specific input segments, thereby enhancing feature representations in various tasks, such as in machine translation and image captioning. The fundamental formula can be denoted as:
\begin{equation}
\text{Attention}(Q, K, V) = \text{softmax}\left(\frac{xQ_w\cdot XK_w^T}{\sqrt{d_k}}\right) XV_w
\end{equation}
where the key-value pairs and queries have different embedding, and $\sqrt{d_k}$ represents the dimension of key which plays a role of scale. There are also several variants of attention mechanisms, including self-attention, multi-head attention, and cross attention, each differing based on the source of the query and the key-value pairs.
Within the specific domain of intelligent Fault Diagnosis (IFD), the introduction of attention mechanisms achieves incredible success\cite{lvAttention2022} in numerous equipment health management tasks, such as fault detection\cite{Beta-VAE}, fault feature enhancement\cite{liu2022data}, anomaly detection\cite{wangWindTurbineFault2023}, remaining useful life prediction\cite{zuoHybridAttentionbasedMultiwavelet2023}, and etc. However, most models use the attention mechanism, especially self-attention, as a way of feature enhancement, while ignoring the possibility of introducing other conditional information through cross-attention.

\subsection {Denoising Diffusion Model}

\begin{figure}[ht!]
\centering
\includegraphics[width=3in]{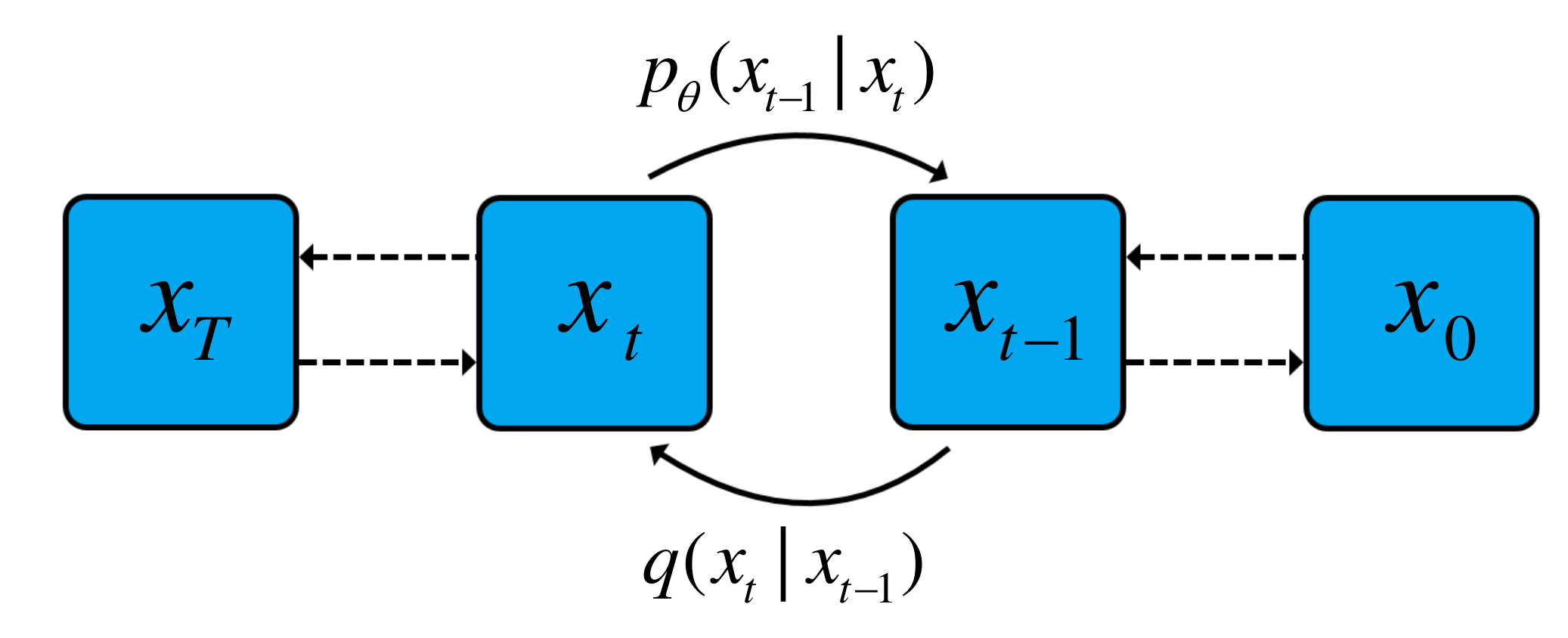}\\
\caption{Process of diffusion model, where $x_0$ is ground truth sample and $x_T$ is nearly Gaussian Noise  }\label{diffuion framework}
\end{figure}

Diffusion models are a family of probabilistic generative models firstly introduced by Sohl et al\cite{sohl2015deep}, which inspired by non-equilibrium thermodynamics. Given a Markov chain of diffusion steps to inject random noise to data and reverse the process to construct nearly ground truth data from the noise. Current research on diffusion models is mostly based on Denoise Diffusion Probabilistic Models (DDPMs)\cite{ho2020denoising}, as shown in Fig.~\ref{diffuion framework}. The overall process gives forward diffusion process $q$ to add small amount of noise to the real sample $x_0$ in $T$ steps and train a gradient neural network $p_\theta$ like U-Net\cite{ronneberger2015unet} to denoise a images from an approximately Gaussian Noise $x_T$. DDPMs have achieved state-of-the-art stable high-quality data generation in the field of Image\cite{song2020score} and audio synthesis\cite{huang2022fastdiff}.  While repetitive denoise steps cause longer training and sampling time, it enables controllability and flexibility to guide the sampling process for improved signals generation quality. 

\section{Proposed Method}

\subsection{Diffusion Model Background}
\begin{figure*}[h]
  \centering
  \includegraphics[width=7in]{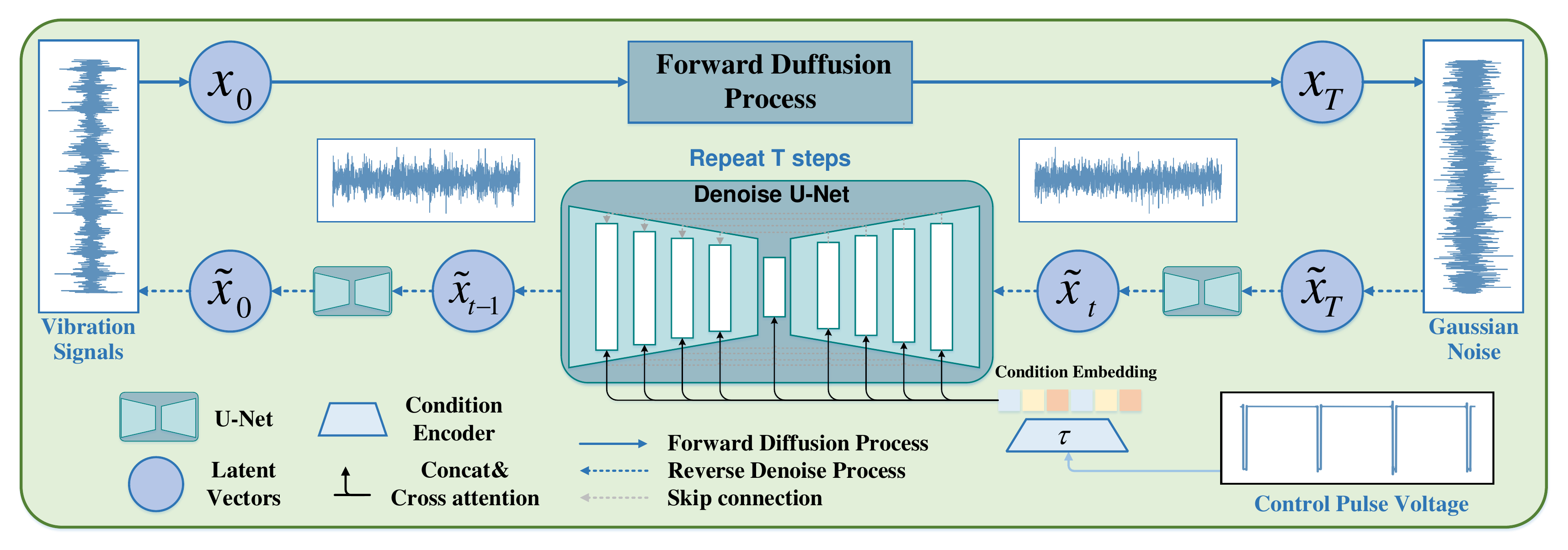}
  \caption{The overall framework of VGCDM, where motor pulse voltage signals are encoded into denoise U-net to guide the reverse process of vibration signals}\label{overall framework}
\end{figure*}
Here we adopt DDPMs architecture improved and validated by Ho et al.\cite{ho2020denoising}. Despite current works are mostly experimented on image data, DDPMs can also generate time series signals like audio or vibration signals with fixed sample points. Hence, given a fixed length signals series $\mathbf{x_0}$ sampled from a real signals distribution $q$ (i.e. $\mathbf{x_0}\sim q_{\mathbf{x_0}}$), the forward process progressively add Gaussian noise and produce a Markov chain of latent variables $\mathbf{x_1},\mathbf{x_2},\ldots,\mathbf{x_T}$:
\begin{equation}
q\left(\mathbf{x}_t \mid \mathbf{x}_{t-1}\right)=\mathcal{N}\left(\mathbf{x}_t ; \boldsymbol{\mu}_t=\sqrt{1-\beta_t} \mathbf{x}_{t-1}, \mathbf{\Sigma}_t=\beta_t \mathbf{I}\right)
\end{equation}
Where $\beta_t \in(0,1)$ is a small hyper-parameter related to time step $t$ to control the ratio of added noise. A common schedule is linear schedule which progressively add the portion of $\beta_t$ as follows: 
\begin{equation}
\beta_t = \beta_{\text{start}} + (\beta_{\text{end}} - \beta_{\text{start}} )\cdot \frac{t}{T - 1}
\end{equation}
Moreover,  due to the magnitude of  $\beta_t$ is tiny, the posterior $q(\mathbf{x_{t-1}}|\mathbf{x_t})$ can be approximately seen as diagonal Gaussian. When taking account of the whole noise added throughout the Markov chain, the total noise is large enough and accordingly the final variable $\mathbf{x_T}$ is close to an isotropic standard Gaussian Distribution. Hence, the whole forward process from $\mathbf{x_0}$ to $\mathbf{x_T}$ can be illustrated as:
\begin{equation}
q\left(\mathbf{x}_{1: T} \mid \mathbf{x}_0\right)=\prod_{t=1}^T q\left(\mathbf{x}_t \mid \mathbf{x}_{t-1}\right)
\end{equation}
Let $\alpha_t = 1 - \beta_t$ and $\bar{\alpha}_t = \prod_{i=1}^t \alpha_i=\prod_{i=1}^t (1-\beta_t)$. When model is more close to a noise after adding amount of Gaussian noise, it can update larger so usually $\beta_1 <\beta_2 < \dots < \beta_T$ and $\bar{\alpha}_1 > \dots > \bar{\alpha}_T$.  In addition, we can sample $t$ steps  $\mathbf{x}_t$ and accordingly $q(\mathbf{x}_t \vert \mathbf{x}_0)$, which can be denoted as : 
\begin{equation}
\begin{aligned} \mathbf{x}_t &= \sqrt{\alpha_t}\mathbf{x}_{t-1} + \sqrt{1 - \alpha_t}\boldsymbol{\epsilon}_{t-1} \\ &= \sqrt{\alpha_t \alpha_{t-1}} \mathbf{x}_{t-2} + \sqrt{1 - \alpha_t \alpha_{t-1}} \bar{\boldsymbol{\epsilon}}_{t-2} & \\ &= \sqrt{\bar{\alpha}_t}\mathbf{x}_0 + \sqrt{1 - \bar{\alpha}_t}\boldsymbol{\epsilon}_{0} \end{aligned}
\end{equation}
\begin{equation}
q(\mathbf{x}_t \vert \mathbf{x}_0)=\mathcal{N}(\mathbf{x}_t; \sqrt{\bar{\alpha}_t} \mathbf{x}_0, (1 - \bar{\alpha}_t)\mathbf{I})  
\end{equation}
Consequently,  whole forward and reverse process is broken into T steps by adding noise. Utilizing a trainable model $p_{\theta}$ to learn to approximate conditional probabilities and run reverse diffusion process. The reverse diffusion process can be denoted as:
\begin{equation}
p_\theta\left(\mathbf{x}_{0: T}\right)=p_\theta\left(\mathbf{x}_T\right) \prod_{t=1}^T p_\theta\left(\mathbf{x}_{t-1} \mid \mathbf{x}_t\right)
\end{equation}
Similarly, a single step reverse process $p_\theta(\mathbf{x}_{t-1}|\mathbf{x}_{t})$ can be estimated by Gaussian Distribution:
\begin{equation}
p_\theta\left(\mathbf{x}_{t-1} \mid \mathbf{x}_t\right)=\mathcal{N}\left(\mathbf{x}_{t-1} ; \boldsymbol{\mu}_\theta\left(x_t, t\right), \Sigma_\theta\left(\mathbf{x}_t, t\right)\right)
\end{equation}
Where starting with a sample noise $\mathbf{\hat{x}_t} \sim \mathcal{N}(0,I)$ and train a learned posterior denoise diffusion model $\epsilon_\theta$ to predict the added noise, then progressively synthesize the sample $\mathbf{x}_0 \sim p_\theta(\mathbf{x}_0)$ at the end. 
Since the variational lower-bound(VLB) used in the DDPM is an effective surrogate objective for the optimization of diffusion model. Simple standard mean-squared error(MSE) to evaluate the loss and  the optimize loss  can be denoted as:
\begin{equation}
L_t = \mathbb{E}_{t \sim [1, T], \mathbf{x}_0, \boldsymbol{\epsilon}_t} \Big[\|\boldsymbol{\epsilon}_t - \boldsymbol{\epsilon}_\theta(\mathbf{x}_t, t)\|^2 \Big]
\end{equation}
This suit more for images-likewise data but not for the synthesize of  vibration due to the possible phase difference of possible signals. Hence, we turn the overall optimization objects to a  Huber loss as:
\begin{equation}
L_t = \mathbb{E}_{t \sim [1, T], \mathbf{x}_0, \boldsymbol{\epsilon}_t} \Big[\text{Huber}(\|\boldsymbol{\epsilon}_t - \boldsymbol{\epsilon}_\theta(\mathbf{x}_t, t)\|)  \Big]
\end{equation}
Where Huber loss is a compromise between MAE and MSE loss, it is smooth around zero but for large errors it behaves more similarly to MAE loss, which can be denoted as :
\begin{equation}
\text{Huber}=
\begin{cases} 
0.5 (\|\boldsymbol{\epsilon}_t - \boldsymbol{\epsilon}_\theta(\mathbf{x}_t, t)\|)^2 & \text{if } \|\boldsymbol{\epsilon}_t - \boldsymbol{\epsilon}_\theta(\mathbf{x}_t, t)\| < 1 \\
\|\boldsymbol{\epsilon}_t - \boldsymbol{\epsilon}_\theta(\mathbf{x}_t, t)\| - 0.5 & \text{otherwise}
\end{cases}
\end{equation}
\subsection{Condition Pulse Voltage Guidance}
\begin{figure}[t!]
\centering
\includegraphics[width=3.2in]{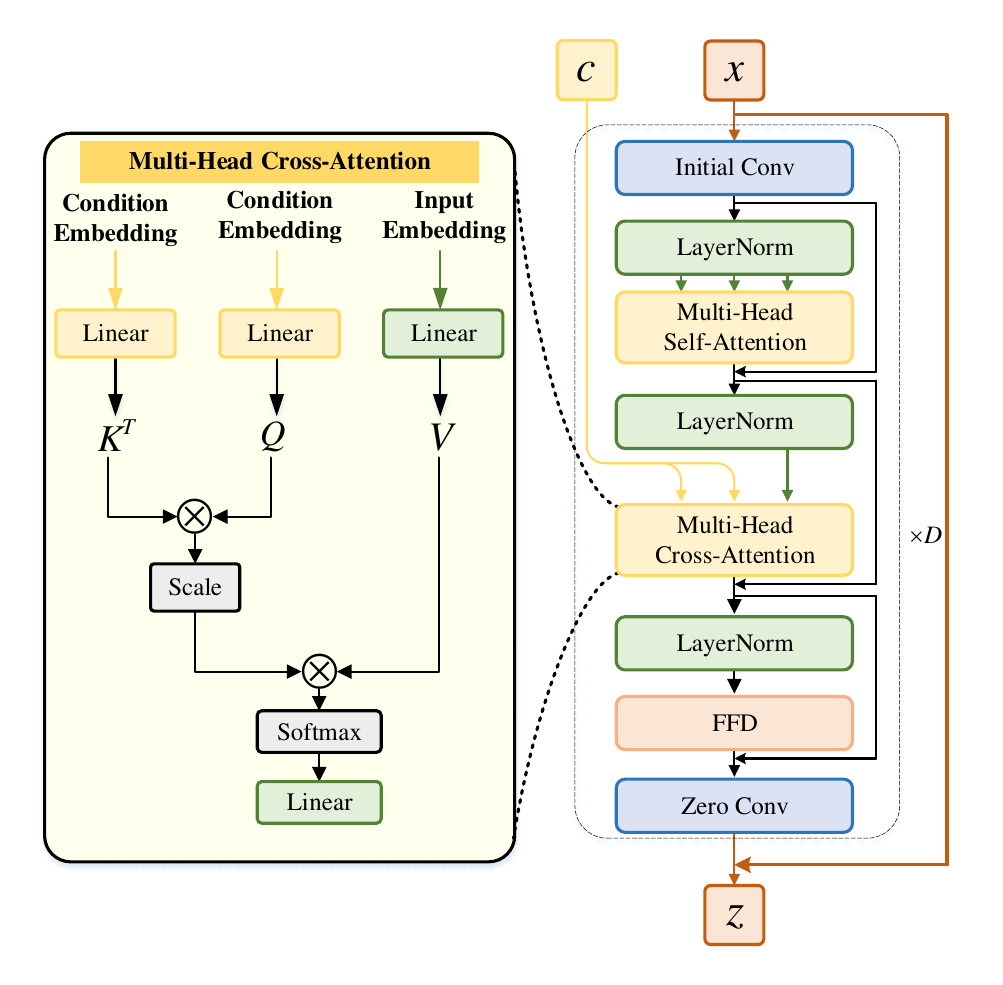}
\caption{Architecture of the Condition Encoder $\tau_\theta$ and Multi-Head Cross-Attention mechanism }\label{cross-att}
\end{figure}

The cross-attention mechanism, a core component of Transformer-based architectures\cite{vaswani2017attention}, has exhibited remarkable proficiency in integrating information from diverse modalities. As illustrated in Fig. \ref{cross-att}, we acquire signals from two different modalities (vibration and voltage) and introduce guidance into the denoising model by computing attention scores using the conditional encoder $\tau_\theta$. Within the depth $D$ of the conditional encoder, the inputs undergo a transformation to an inner dimension through an initial convolution. Each transformation is followed by Layer Normalization \cite{ba2016layer} to maintain consistent scaling information. Prior to computing the cross-attention score, input embedding are processed with multi-head self-attention to extract latent periodic features.
In the cross-attention operation, the attention score can be formulated as:
\begin{equation}
\text{Attention}(Q, K, V) = \text{softmax}\left(\frac{QK^T}{\sqrt{d_k}}\right) V
\end{equation}
Here, the keys and queries are derived from the conditional control information $c$, specifically the motor pulse voltage signals in following cases. Their values are obtained from the latent embedding $h_i$ of input signal and conditions signal $c$, it can be denotes as:
\begin{equation}
 \begin{aligned}Q = W^{(i)}_Q \cdot f(h_i),\; K = W^{(i)}_K \cdot g(c),\; V = W^{(i)}_V \cdot v(c) \end{aligned}
\end{equation}
Where $f(\cdot),g(\cdot),v(\cdot)$ is Linear function to map the input to the same dimension. Following a standard feed-forward operation, the embedding are processed with a zero convolution. This implies that the initial weights and biases are set to zero to prevent the control signals from influencing the initial training of the denoising model. Subsequently, the conditional attention score can be added to the input to get the implicit output $z$ by residual connection, which can be denoted as:
\begin{equation}
    z=x+\tau_\theta(x,c)
\end{equation}

Architecture of the condition encoder and  the overall framework is shown in Fig\ref{overall framework}
Accordingly,  condition $c$ can be added in the loss function, which can be expressed as:
\begin{equation}
L_t= \mathbb{E}_{t \sim [1, T], \mathbf{x}_0, \boldsymbol{\epsilon}_t} \Big[\text{Huber}(\|\boldsymbol{\epsilon}_t - \boldsymbol{\epsilon}_\theta(\mathbf{x}_t, t,c)\| )\Big]
\end{equation}
Overall algorithms can be summarized in Algorithms.\ref{alg1} and Algorithms.\ref{alg2}. In the training process, we sample the vibration signals and accompanied conditional control signals. Aligning input embedding with condition embedding until the denoise network $\theta$ converge. In the sampling process, input is sorely condition signal $c$.  Sample vibration signal shape likewise noise and denoise noise by pre-trained denoise model $\theta$ to recover a high quality vibration signals.
\begin{algorithm}[t!] 
	\caption{Training Denoise U-net $\theta$}
	\label{alg1} 
	\begin{algorithmic}[1]
	   \REPEAT
    	\STATE $(\mathbf{x_0},c)\sim q(\mathbf{x_0},c)$
    	\STATE $t\sim Uniform(\{1,\dots,T\})$
    	\STATE $\boldsymbol{\epsilon}\sim\mathcal{N}(0,1)$ 
          \STATE Take gradient desent step on \\ \hspace{1cm}$\nabla_{\theta}\|\boldsymbol{\epsilon} - \boldsymbol{\epsilon}_\theta(\sqrt{\bar{\alpha}_t} \mathbf{x}_0+\sqrt{1-\bar{\alpha}_t} \boldsymbol{\epsilon}, c,t)\|^2 $
	   \UNTIL converged 
	\end{algorithmic} 
\end{algorithm}
\begin{algorithm}[t!] 
	\caption{Sampling}
        \hspace*{0.02in} {\bf Input:} condition signal $c$ and Unet parameters $\theta$\\
        \hspace*{0.02in} {\bf Output:}
        generated signal $\mathbf{\hat{x}_0}$
	\label{alg2} 
	\begin{algorithmic}[1]
    	\FOR{$t=T,\dots,1$} 
    	   \STATE $\mathbf{\hat{x}}_{t} \sim\mathcal{N}(0,1)$ 
            \STATE Sample $\mathbf{\hat{x}_{t-1}} \sim p_{\theta}(\mathbf{\hat{x}_{t-1}}|\mathbf{\hat{x}_{t};c;t})$
            \ENDFOR
	\end{algorithmic} 
\end{algorithm}

\section{Cases study}

To prove experimentally the effectiveness of proposed method, in this section, two cases utilizing dataset respectively collected from Spectra Quest(SQ)  and HSR bogie Bearing Platform are tested. 

\subsection {Comparison Methods and Metrics}
We provide a concise overview of four generative model-based algorithms for signal generation, setting the stage for comparative analysis. Following this, we identify three metric indexes to evaluate and contrast the outcomes generated by these models.

\begin{enumerate}
\item VQVAE: A Vector Quantized-VAE variant that introduces a discrete latent space using vector quantization, enhancing the model's ability to capture more complex data distributions.
\item DCGAN: Deep Convolutional Generative Adversarial Network employs deconvolutional layers, enabling the generation of high-resolution data with intricate details from random noise.
\item QSCGAN: Quick Self-attention Convolution GAN utilizes self-attention and spectral normalization to ensure the stability and convergence rate of output quality and input similarity.
\item DDPM: Denoising Diffusion Probabilistic Models use stochastic differential equations to drive the data distribution to the model's prior, comparing the proposed model lack of the cross attention guidance.
\end{enumerate}

Comparative methods have adopted suitable experimental configurations. For VAEs-based and  GAN-based methods, we employed 3000 training epochs to achieve full convergence. In the DDPMs-based methods, the training process consists of 200 epochs, and $\beta_t$ is linearly decayed over a total of T = 1000 time steps, with $\beta_{start}$ = 0.0001 and $\beta_{end}$ = 0.02. A single vibration signal sample, along with the motor pulse voltage signal from dataset, contains 2048 data points with a normalization to a range of -1 to 1. All data has been sliced into non-overlapping samples. 70\% of these samples are designated for the training dataset, while the remaining 30\% are used as the test dataset. The chosen optimizer is Adam\cite{kingma2014adam}, initialized with a learning rate of 0.0001 and a weight decay of 0.1. 

To verify the generation performance of proposed methods under different fault condition, We construct three indexes to evaluate the performance of generation quality: Root-Mean-Square Error(RMSE), Peak Signal-to-Noise Ratio(PSNR) and Frequency Spectrum Cosine Similarity(FSCS). Here, $y_i$ refers to the ground-truth sample and $\hat y_i$ is the sample generated based on the accompanying conditional impulse voltage signal as well as the noise.

RSME refers to the quadratic mean of the differences between the actual signal and the generated signal in the time domain space. In general, a lower RMSE indicates a better generated quality.
\begin{equation} 
\text{RMSE}(Y,\hat{Y}) = \sqrt{\frac{1}{n} \sum_{i=1}^{n}(y_i-\hat{y}_i)^2}
\end{equation}

PSNR quantifies the relationship between the highest achievable signal power and the power of disruptive noise that impacts the accuracy of its representation. A higher PSNR generally represents a better reconstruction quality.
\begin{equation}
\text{PSNR}(Y,\hat{Y}) = \frac{1}{n} \sum_{i=1}^{n}\ \left(10 \times \log_{10}\left( \frac{\text{MAX}_{y_i}^2}{\text{MSE}(y_i, \hat{y}_i)}\right)\right)
\end{equation}

Although these two metrics reflect the quality of the generated data to some extent , possible phase differences in the time-domain signal generation may lead to distortions in the quantization results. Since the generated data and the actual data have the same sampling point and sampling frequency, their scales on the frequency spectrum are the same, so here we utilize the $\text{FSCS}\in(0,1)$ to evaluate the spectral similarity, And a higher FSCS indicate higher quality of the generated signals.
\begin{equation}
\text{FSCS}(Y,\hat{Y}) = \frac{1}{n} \sum_{i=1}^{n}\frac{fft(y_i) \cdot fft(\hat{y}_i)}{\| fft(y_i) \| \| fft(\hat{y}_i) \|}
\end{equation}

\subsection {Case Study1: Spectral Quest Dataset}
\begin{figure}[t!]
\centering
\includegraphics[width=3.0in]{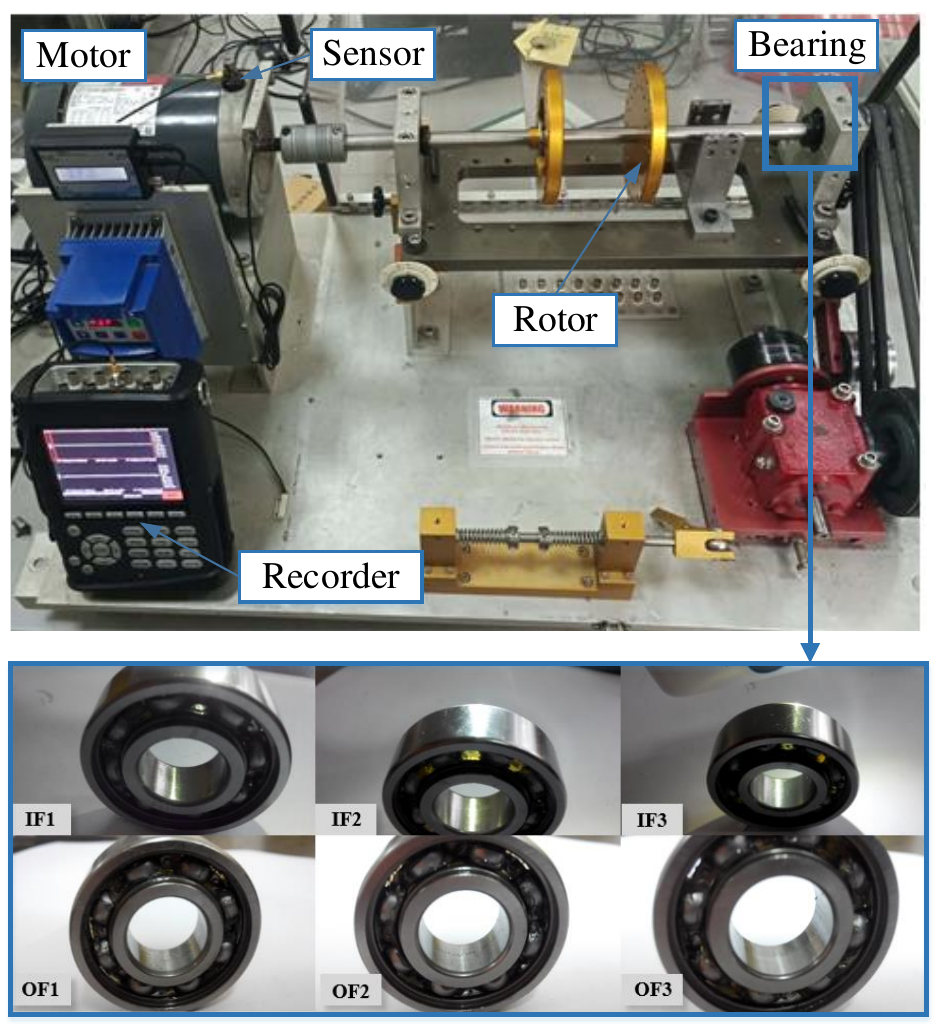}
\caption{Top is SQ test rig. Bottom is six single point fault bearings with different fault location and degree of failure}\label{SQV test rig}
\end{figure}

\begin{table*}[ht!]
    \caption{SIGNAL GENERATION QUALITY USING SQ EXPERIMENTAL UNDER OF3 CONDITION AND 19Hz ROTATION FREQUENCY}
    \centering
    \begin{tabular}{l c c c c}
    \toprule
    Methods & Sample or Guidance & \hspace{1cm} RSME$\downarrow$ \hspace{1cm} & \hspace{1cm} PSNR$\uparrow$ \hspace{1cm} & 
    FSCS$\uparrow$ \hspace{1cm}\\
    \midrule
    VQVAE & samples & 0.8410$\pm$0.0043 & 1.5307$\pm$0.4606 & 0.1977$\pm$0.0153 \\
    DCGAN & noise & 0.4155$\pm$0.0021 & 7.6806$\pm$0.8866 & 0.5426$\pm$0.0023 \\
    QSCGAN & noise & 0.4265$\pm$0.0004 & 7.7387$\pm$0.2715 & 0.5504$\pm$0.0024 \\
    DDPM & noise & 0.3875$\pm$0.0003 & 8.2415$\pm$0.1407 & 0.6925$\pm$0.0005 \\
    \textbf{VG-CDM} & (noise,c) & \textbf{0.3770}$\pm$\textbf{0.0002} & \textbf{8.4813}$\pm$\textbf{0.1145} & \textbf{0.7354}$\pm$\textbf{0.0006} \\
    \bottomrule
    \end{tabular}
    \label{tab:sq_result1}
\end{table*}

\begin{figure}[ht!]
\centering
\includegraphics[width=3.2in]{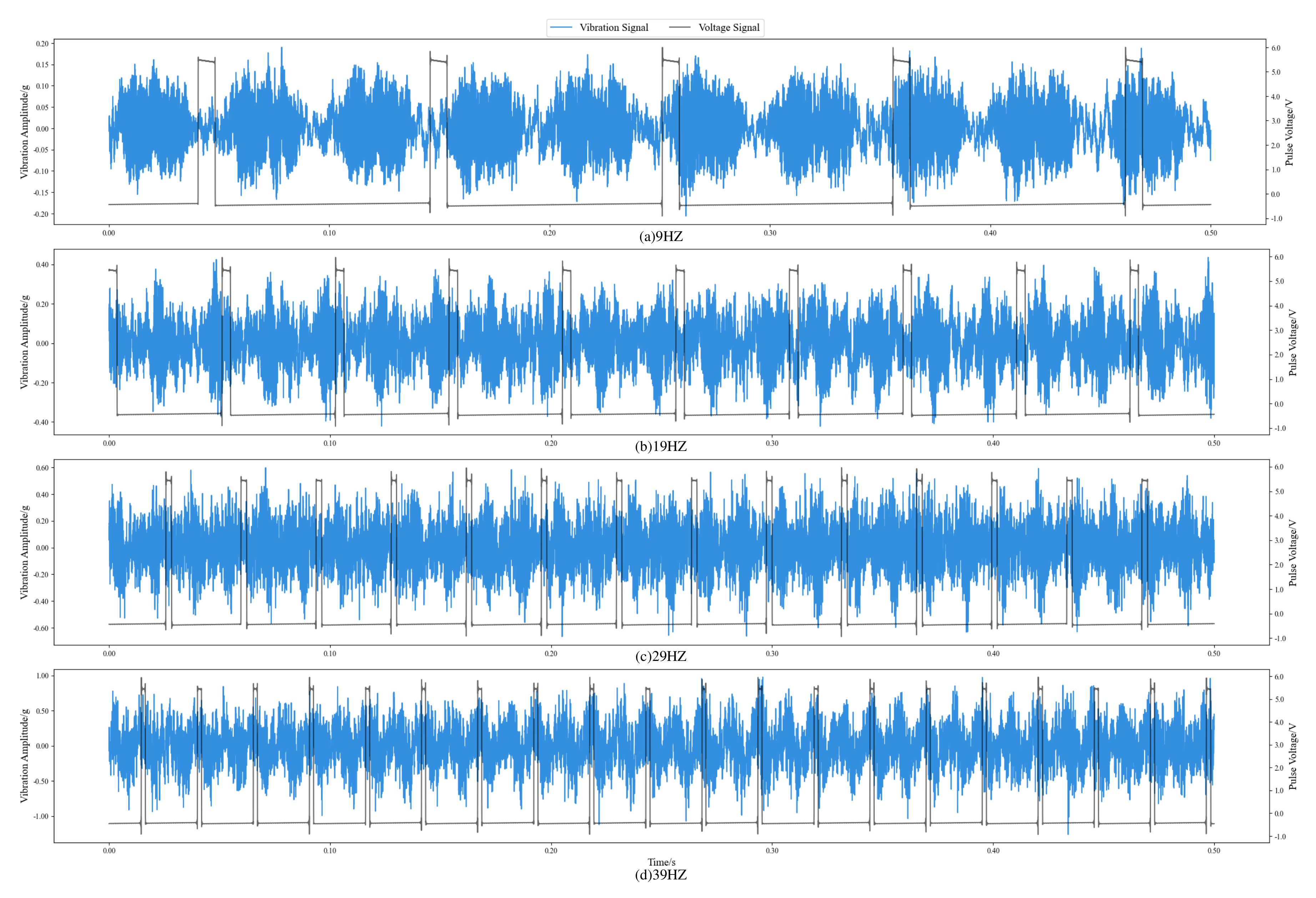}
\caption{Vibration (in blue) and Voltage (in black) signals under the NC condition for four steady-state rotating frequencies (9Hz, 19Hz, 29Hz, 39Hz), all with the same half-second sample duration. Higher rotating frequencies mean denser vibration signals and pulsed voltage signals.}\label{fault-bears}
\end{figure}

\subsubsection{Steady-State Condition}
 As Fig. ~\ref{fault-bears} shows, collected vibration data contain four steady motor rotating frequency: 9Hz, 19Hz, 29Hz and 39 Hz. Higher rotating frequency indicate denser vibration signals and pulsed voltage signals. Firstly, we conduct an experiment of different methods in 19Hz and OF3 condition. The compared results are demonstrated in Table \ref{tab:sq_result1}, our proposed method achieve the best performance in RSME, PSNR and FSCS, which indicates that the proposed methods have better generated similarity under a single steady working state. Simultaneously, the results demonstrate minimal variance, indicating enhanced stability in the outputs generated by the proposed method.
\begin{table}[ht!]
    \centering
    \caption{SIGNALS GENERATION QUALITY USING SQ EXPERIMENTAL UNDER STEADY-STATE CONDITION}
    \begin{tabular}{c c c c}
    \hline\hline
    Conditions&  RSME$\downarrow $& PSNR $\uparrow$& FSCS$ \uparrow$\\
    \hline
        NC& 0.3953$\pm$0.0012&	8.0931$\pm$0.5306&	0.6945	$\pm$0.0070 \\
        IF1&  \textbf{0.2867}$\pm$\textbf{0.0006}&	\textbf{10.8854}$\pm$\textbf{0.5739}	&0.7083	$\pm$0.0023 \\
        IF2 &  0.2993$\pm$0.0007&	10.5088$\pm$0.5460&	0.6896	$\pm$0.0038 \\
        IF3 &  0.3460$\pm$0.0006&	9.2389$\pm$0.3546&	0.7129	$\pm$0.0018 \\
        OF1 &  0.3432$\pm$0.0009&	9.3227$\pm$0.5650&	0.7241	$\pm$0.0019\\
        OF2 &  0.3963$\pm$0.0005&	8.0554$\pm$0.2594&	0.7271	$\pm$0.0012 \\
        OF3 &  0.3895$\pm$0.0004&	8.2011$\pm$0.1868&	\textbf{0.7385}	$\pm$\textbf{0.0013} \\
    \hline\hline
    \end{tabular}
    \label{tab:sq_result2}
\end{table}

To delve deeper into the influence of various steady rotation frequencies on signal generation, we executed an experiment where training samples from all four steady-state speed were amalgamated. As depicted in Table \ref{tab:sq_result2}, the mixed OF3 condition attained an FSCS of 0.7385, showing no difference compared to operations at 19Hz. Moreover, generative models trained across different bearings demonstrated comparable generative capabilities, indicating a uniformity in performance across the diverse conditions.

\subsubsection{Varying-State Conditions} 
\begin{figure}[ht!]
\centering
\includegraphics[width=3.2in]{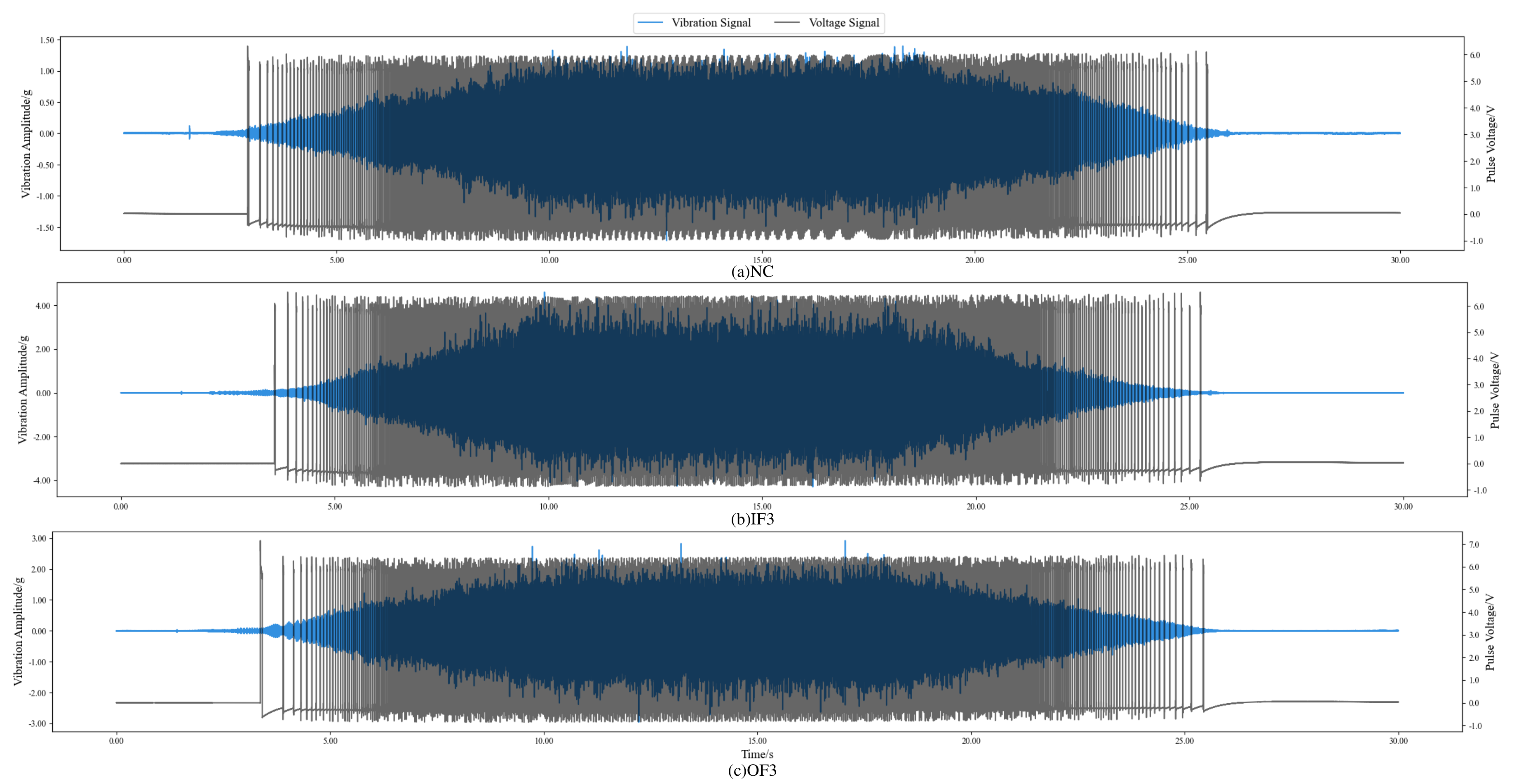}
\caption{A complete sampling of Vibration (in blue) and Voltage (in black) signals under SQ vary-state condition. The control signals for the various speeds are essentially the same for all different bearings (here NC, IF3, OF3). Interval of the control pulse voltage indirectly reflects the speed change of the motor}\label{SQV-fig}
\end{figure}

We further explore the performance under a more general yet challenging varying state scenario. As depicted in Fig. \ref{SQV-fig}, we implement a similar control process across different bearings: initially standstill,  acceleration, steady speed operation, deceleration, and finally returning to standstill. Throughout this process, the control pulse voltage indirectly reflects the motor's speed changes. Similarly with experiment in steady-state conditions, we slice sample with a fixed length of 2048, with most samples containing either no pulse (indicating the rest state) or 3 to 4 pulse signals (representing the steady state). The presence of 1 or 2 pulses indicated an acceleration or deceleration process. 

Due to the adverse impact of varying-state conditions on the performance of the comparison methods without condition guidance, the value of the comparison diminishes. Hence, our primary focus is on assessing the generated stability of different bearing conditions and pulse intervals on signal generation.  Table \ref{tab:sqv_result} deceits the generated results for seven different bearings. Compared to steady-state experiments,  performance does experience a reduction but still maintains an FSCS result above 0.6 in all seven cases. 

\begin{figure*}[bt!]
  \centering
  \includegraphics[width=7in]{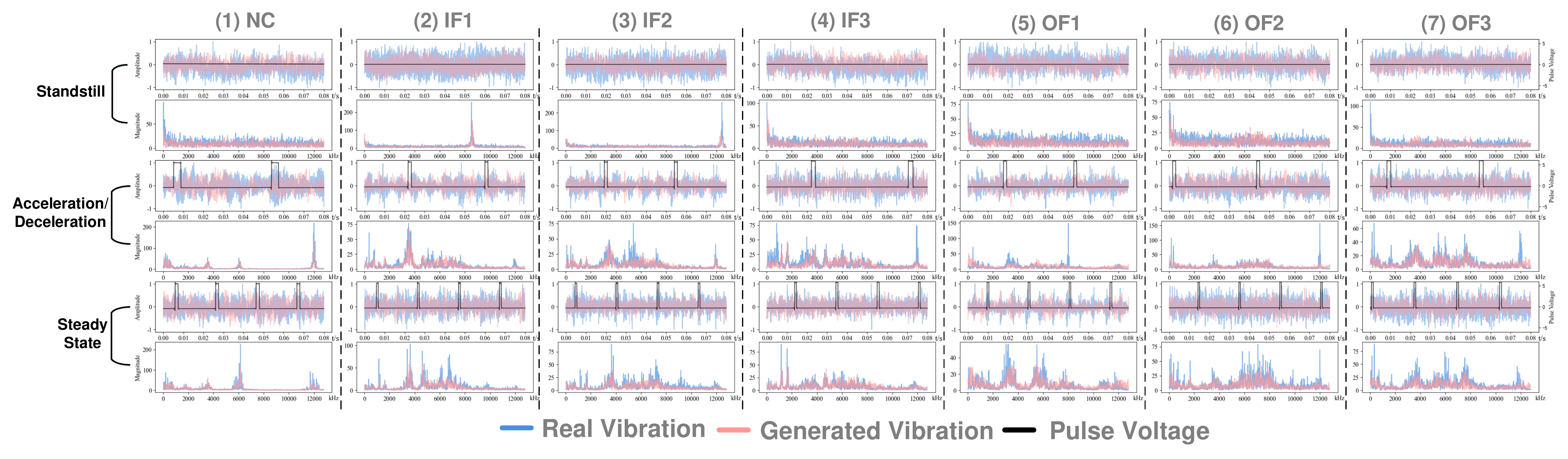}
  \caption{ Visualizations comparison of vibration and control voltage signals across three primary operational states (standstill, acceleration/deceleration, and steady state) in Seven Bearings.  Generated vibration signals are solely derived from the sampling real control voltage. Each graph is structured with upper and lower panels. The upper panel presents a comparison in time domain, while the lower panel displays a frequency spectrum comparison.}\label{SQV_Compare}
\end{figure*}

Moreover, we conduct visualizations of both real data and generated data alongside control pulse voltage signals of seven various bearings experiments across three primary operating stages. As presented in Fig. \ref{SQV_Compare}, the model exhibits improved frequency spectrum reconstruction representation performance in a relative steady-state process such as standstill or steady operation. For acceleration and deceleration phases, the high-frequency components of reconstruction performance of OF condition tend to be less satisfactory which may attribute to fewer pulsed voltage-vibration pair samples are available or differences between the acceleration and deceleration. Despite these challenges, the generated vibration signals aligned well with the control voltage signals, effectively capturing feature information.
\begin{table}[ht!]
    \centering
    \caption{SEVEN BEARING SIGNALS GENERATION METRICS USING SQ EXPERIMENTAL UNDER VARY-STATE CONDITION}
    \begin{tabular}{c c c c}
    \hline\hline
    Conditions&  RSME$\downarrow $& PSNR $\uparrow$& FSCS$ \uparrow$\\
    \hline
        NC& 0.4036$\pm$0.0011&	7.9115$\pm$0.5382&	0.6070$\pm$0.0183 \\
        IF1&  0.3739$\pm$0.0040&	8.6641$\pm$2.0625&	0.6149$\pm$0.0240 \\
        IF2 &  \textbf{0.3347}$\pm$\textbf{0.0027}&	\textbf{9.6038}$\pm$\textbf{1.6530}&	0.6095$\pm$0.0165 \\
        IF3 &  0.3674$\pm$0.0018&	8.7493$\pm$0.8763&	0.6524$\pm$0.0207 \\
        OF1 &  0.3428$\pm$0.0023&	9.3792$\pm$1.3761&	0.6548$\pm$0.0181\\
        OF2 &  0.3883$\pm$0.0011&	8.2448$\pm$0.4726&	\textbf{0.6688}$\pm$\textbf{0.0181} \\
        OF3 &  0.3859$\pm$0.0012&	8.3021$\pm$0.5210&	0.6589$\pm$0.0197 \\
    \hline\hline
    \end{tabular}
    \label{tab:sqv_result}
\end{table}

\subsection {Case Study 2: High-Speed Rail Bogie Bearing Dataset}
\subsubsection{Experiments Setup}
\begin{figure}[t!]
\centering
\includegraphics[width=3.5in]{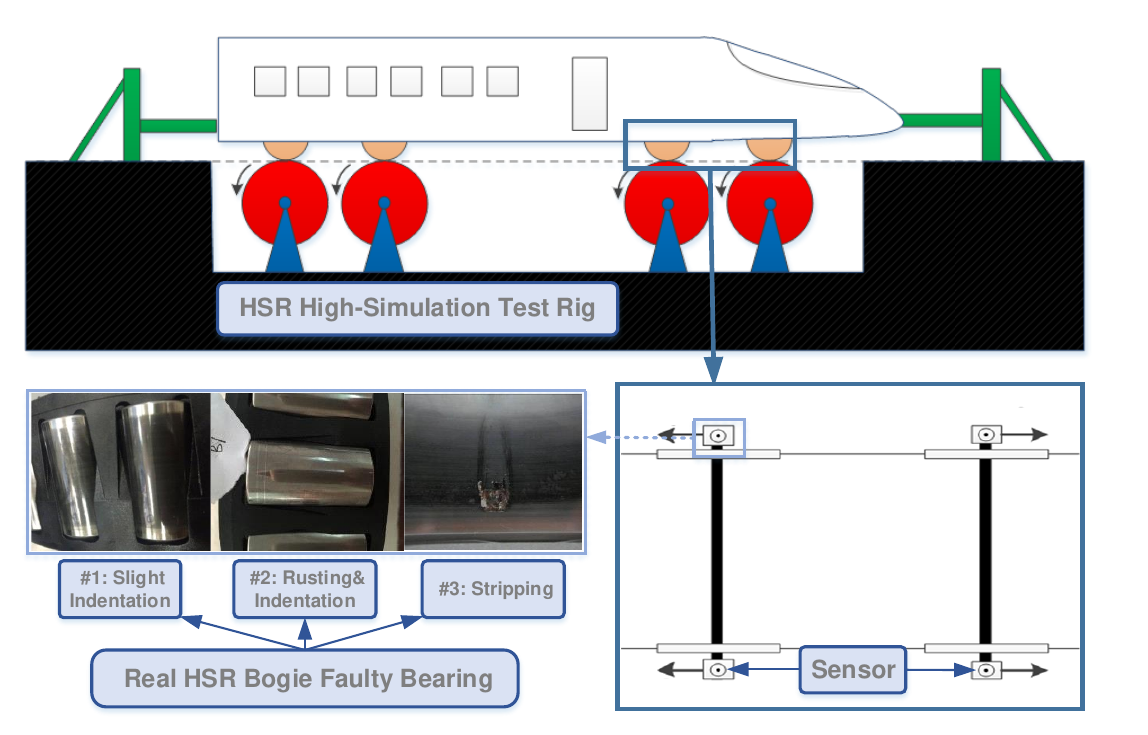}
\caption{The HSR dynamical high-simulation test rig. Sensors are placed on the top of bogie bearing. Replace real bogie faulty bearings to simulate the actual faulty vibration signals}\label{HS test rig}
\end{figure}

To substantiate the efficacy of the proposed method within authentic operational contexts of HSR, we conduct experiments utilizing a high-simulation HSR test rig. The rig replicate the dynamics working condition of a HSR train under controlled voltage conditions. The schematic representation of the experimental setup is illustrated in Fig. ~\ref{HS test rig}. The experimentation involved rolling trial that an train assembly to replicate real scenarios. Vibration sensors were placed on the vertical axis of the bogie’s front dual axles, ensuring precision in data capture. We replaced test rig bogie bearings with actual faulty bearings, subjected to various predefined conditions to simulate vibration. These bearings states included normal functioning (\#0), minor indentation (\#1), concurrent rusting and indentation (\#2), and material stripping (\#3). The sensitivity of sensors is 50 mV/g and a 25.6 kHz sampling frequency for data recorder. The type of test bearings is NTNCRI-2692. Table \ref{tab:hsr_data} is the test condition, which contains uniform acceleration and steady-state operation process close to real HSR working condition. Test speed locate in a range from 100km/h to 350km/h. We sample 120 samples for every acceleration process and steady-state process. Samples are divided into 70\% for training and 30\% for testing. Other experiment setting is same as case1.
\begin{table}[ht!]
    \caption{HSR EXPERIMENTAL TEST CONDITIONS}
    \centering
    \begin{tabular}[width=3.5in]{c c c}
    \toprule
    No. & Last Time & working conditions\\
    \midrule
    0 &  0-20s & uniform acceleration from 100km/h to 200km/h\\
    1 &  20-140s& steady speed lasts for 120s at 200km/h\\
    2 &  140-160s& uniform acceleration from 200km/h to 250km/h\\
    3 &  160-280s& steady speed lasts for 120s at 250km/h\\
    4 &  280-300s& uniform acceleration from 250km/h to 300km/h\\
    5 &  300-420s& steady speed lasts for 120s at 300km/h\\
    6 &  420-440s& uniform acceleration from 300km/h to 350km/h\\
    7 &  440-560s& steady speed lasts for 120s at 350km/h\\
    \bottomrule
    \end{tabular}
    \label{tab:hsr_data}
\end{table}
\subsubsection{Results and Analysis}
\begin{figure}[ht!]
\centering
\includegraphics[width=3.2in]{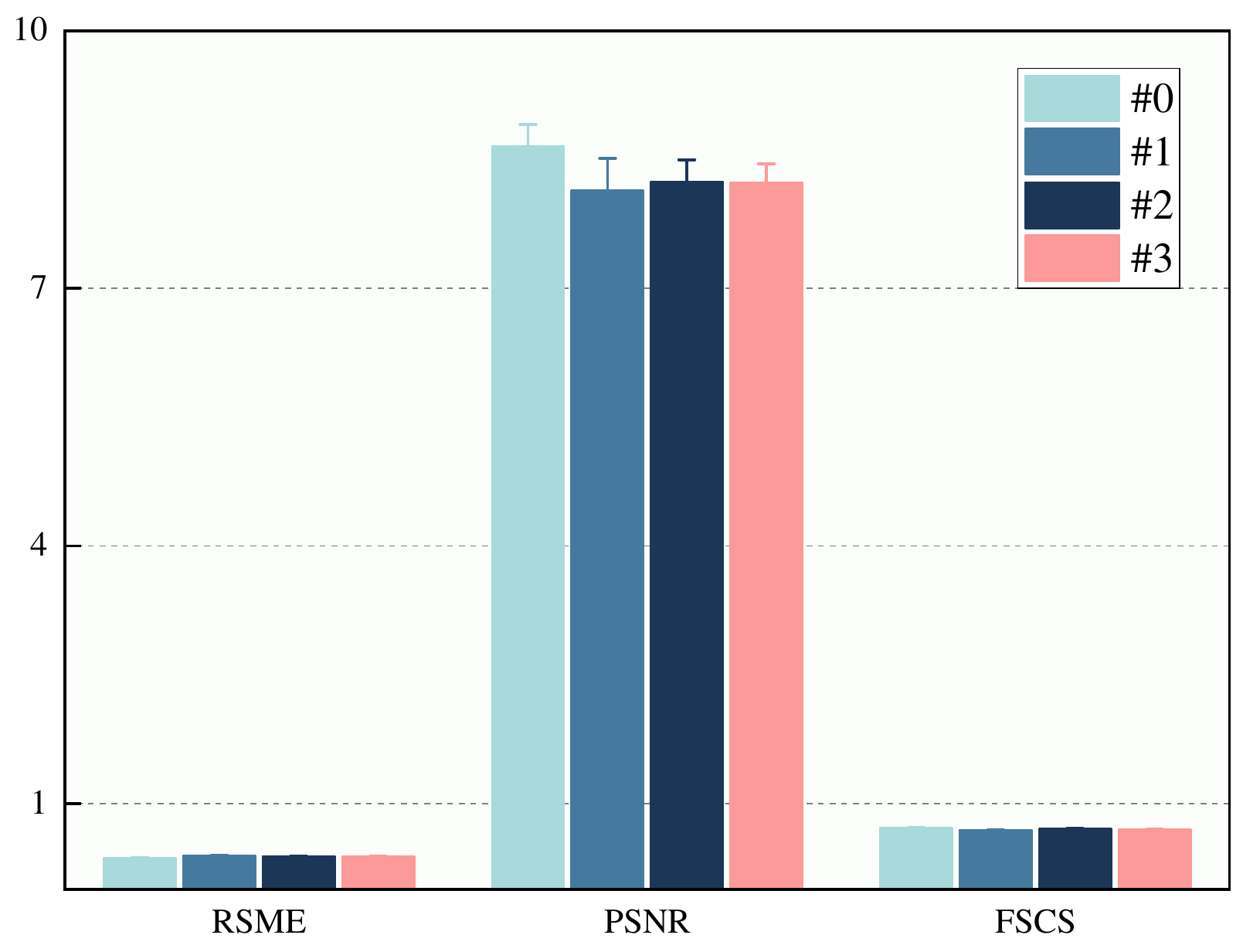}
\caption{Metric Indexes for four different bearings}\label{hs_exp}
\end{figure}

\begin{figure}[ht!]
  \centering
  \includegraphics[width=3.5in]{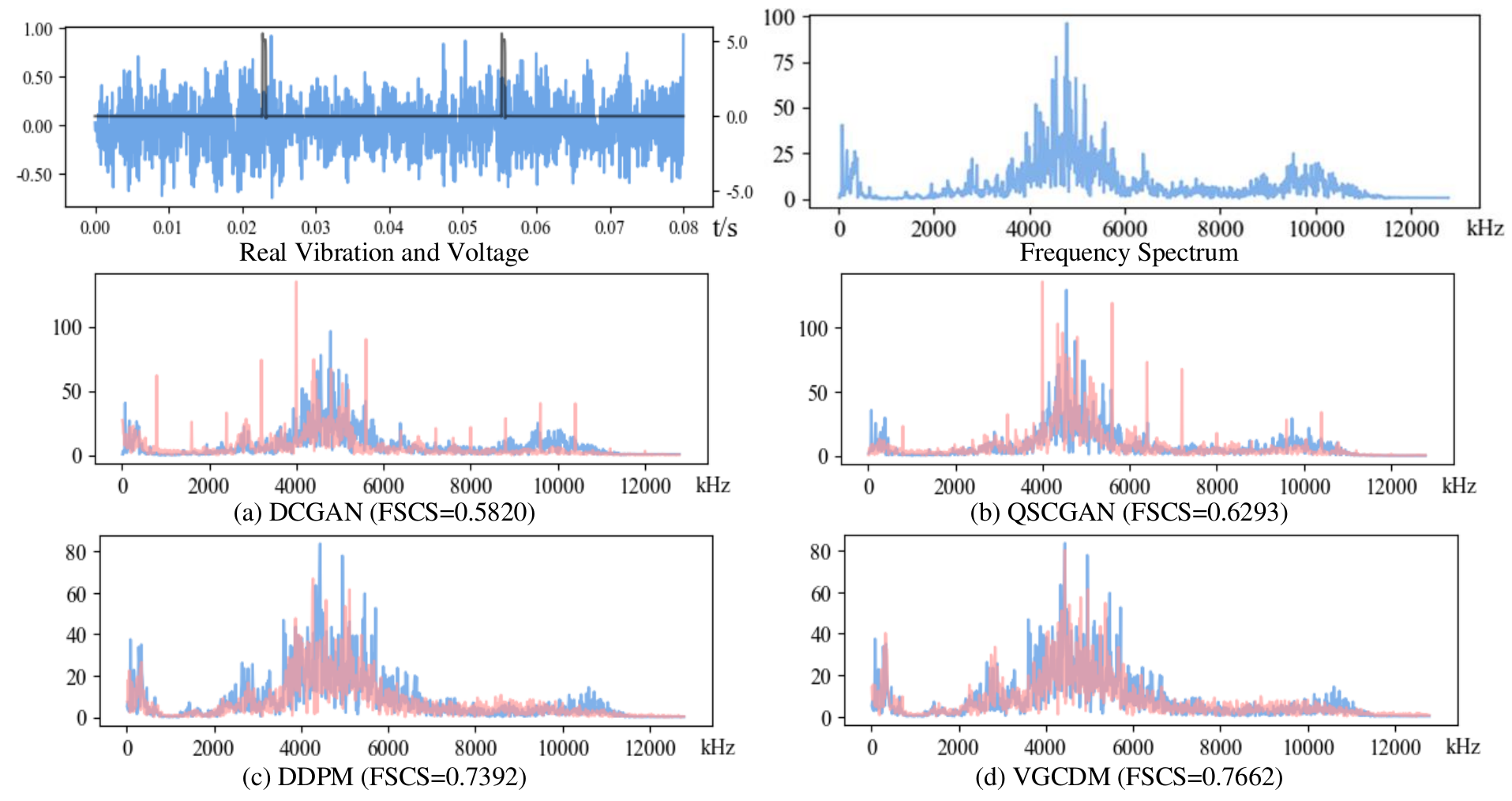}
  \caption{The upper left graph shows a sampled real test vibration signal and the corresponding control voltage signal. The upper right graph is the frequency spectrum. Following graphs show the comparison between the generated(in blue) and real(in red) spectrum, and FSCS of four different models}
  \label{vis of hs}
\end{figure}

Table. \ref{tab:hsr_result1} depicts the compared metrics performance of comparison methods in a normal working conditions. The proposed method still achieve over PSNR of 8.5 and FSCS of 0.7, which have the best reconstructed performance close to laboratory operating environments.  As Fig. \ref{hs_exp} shows the generated performance does not show a difference for four different faulty bearings which keep a similar control process and sample, thereby the generation is stable and close to real operation. 
\begin{table}[ht!]
    \caption{ COMPARISONS PERFORMANCE FOR HSR BOGIE BEARING EXPERIMENTS UNDER NORMAL CONDITION}
    \centering
    \begin{tabular}{l c c c}
    \toprule
    Methods &  RSME$\downarrow$ &PSNR$\uparrow$ & 
    FSCS$\uparrow$\\
    \midrule
    VQVAE &  0.5321$\pm$0.0022 & 5.5144$\pm$0.6172 & 0.2587$\pm$0.0121 \\
    DCGAN &  0.3897$\pm$0.0004 & 8.1959$\pm$0.2056 & 0.5986$\pm$0.0016 \\
    QSCGAN &  0.4269$\pm$0.0018 & 7.4330$\pm$0.6439& 0.6061$\pm$0.0008 \\
    DDPM &  0.3879$\pm$0.0007 & 8.2372$\pm$0.2095 & 0.6822$\pm$0.0020 \\
    \textbf{VGCDM} & \textbf{0.3701}$\pm$\textbf{0.0004} & \textbf{8.6560}$\pm$\textbf{0.2495} & \textbf{0.7173}$\pm$\textbf{0.0025} \\
    \bottomrule
    \end{tabular}
    \label{tab:hsr_result1}
\end{table}
We also visualize a common 250km/h steady-state condition sample as shown in Fig. \ref{vis of hs}. The upper graph demonstrates a sampled real test vibration signal, corresponding control voltage signal and frequency spectrum. The following shows the comparison of spectrum visualization between the GANs-based  and the DMs-based approaches. GAN-based methods will reconstruct some frequency components that are not existed, while DM-based methods will reconstruct closer to the average frequency component components. In addition, voltage guidance help achieve more accurate and detailed generation with a FSCS improved from 0.7392 to 0.7662. 

\section{Discussion}
\subsection{The Visualization of Guidance Attention}
In our proposed method, we align the generated signals with pulse voltage control signals and introduce guidance through a cross-attention mechanism. This cross-attention mechanism plays a crucial role in controlling the denoising process. To gain insight into the impact of the cross-attention mechanism under different conditions, we present multi-channel attention scores for the final time steps of four typical pulse voltage conditions. These scores are derived from a pre-trained model operating under normal varying-state working conditions of SQ dataset, which represents the most common scenario. The attention scores are demonstrated in $\mathbb{R}^{32\times2048}$  format and illustrate in heat map.
\begin{figure}[ht!]
  \centering
  \includegraphics[width=3.5in]{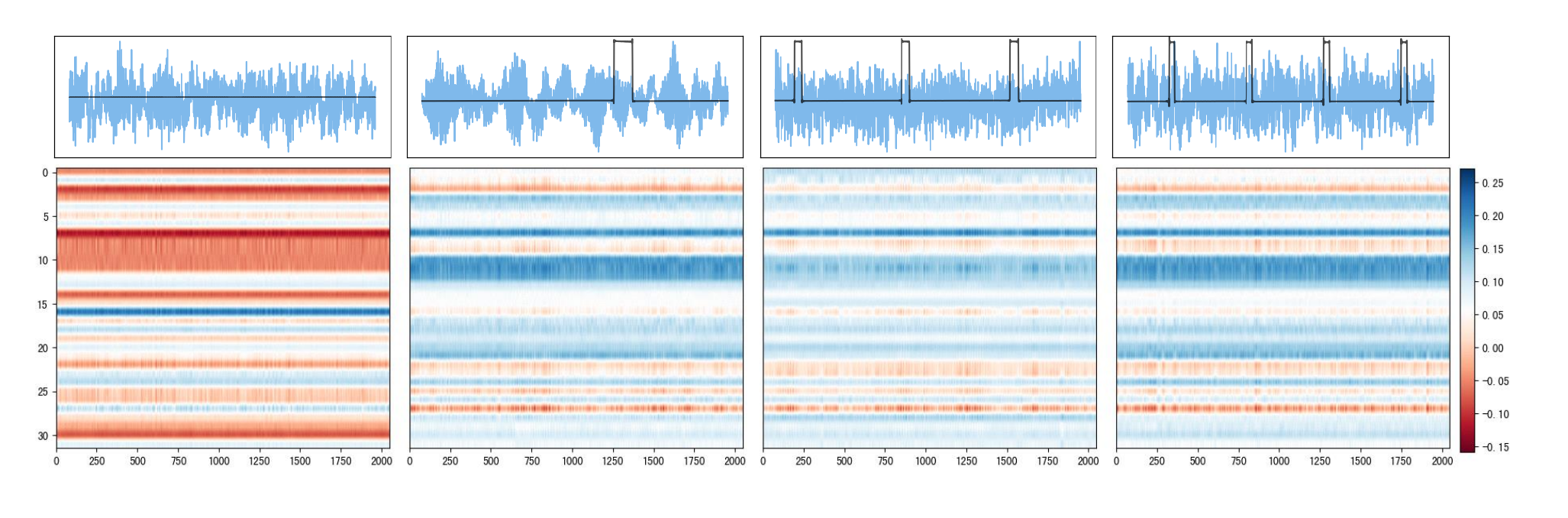}
  \caption{Visualizations of the final 32-channel attention scores across four distinct operating stages. The top panel displays the attention scores, while the bottom panel presents the real data alongside the corresponding control voltage.}
  \label{visualization of attention}
\end{figure}

As illustrated in Fig. \ref{visualization of attention}, the first condition corresponds to a scenario that is no control signal, indicating that the motor is under standstill and vibration signals resemble a Gaussian Noise. In this condition, attention scores exhibit predominantly negative values across most channels that different from other working patterns. Hence, model emphasize less denoising, resulting in generated signals resembling noise. The next three conditions pertain to different speeds, with the third and fourth having the more samples number from a steady working state, while the second condition has the fewest samples from variable-speed conditions. The third and fourth conditions, with their larger sample sizes, yield attention scores that closely align with the actual control voltage description. Conversely, the second condition, characterized by a relatively smaller sample size, exhibits attention scores that close to the fourth condition,  which may resulting in less accurate reconstructed results. In summary, the proposed model have ability to leverage voltage features and incorporate conditional information to generate vibration signals that align with different control signals.
\subsection{Hyper-parameters of diffusion models}
\begin{figure}[ht!]
\centering
\includegraphics[width=3.2in]{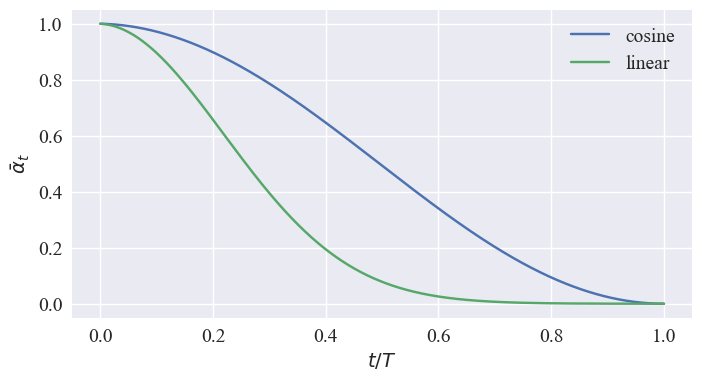}
\caption{The proportion of accumulated noise, denoted as $\bar{\alpha}_t$, is compared between linear and cosine-based schedules of $\beta_t$ in the reverse process}\label{beta_t}
\end{figure}
\subsubsection{$\beta_t$ Schedules}
Given that the diffusion forward process introduces noise to the signals and the denoising model learns to reverse this process, the ratio of noise added at each step in the forward process becomes a critical factor. The choice of $\beta_t$ schedules might  influence the generated performance. Hence, we primarily compare two main scenarios: linear schedules and cosine schedules. We use a comparison of $\bar{\alpha}_t$ as depicted in Fig. \ref{beta_t}, which represents the proportion of accumulated noise that remains for the reverse denoising process. Metrics indexes under normal condition and SQ vary-state experiment are shown in Table\ref{tab:dis_2}. It's evident that that both exhibit comparable performance when used on different schedules. Cosine schedules excel in RSME and PSNR metrics, whereas linear schedules shine in FSCS. Overall, $\beta_t$ Schedules doesn't play a dominant role. 

\begin{table}[ht!]
    \centering
    \caption{PERFORMANCE METRICS USING DIFFERENT $\beta_t$ SCHEDULES UNDER NC AND VARY-STATE CONDITION}
    \begin{tabular}{c c c}
    \hline\hline
    Metric Indexes&Linear&Cosine\\
    \hline
    RSME$\downarrow$ & $0.4036\pm0.0011$& $\textbf{0.3739}\pm\textbf{0.0040}$\\
    PSNR$\uparrow$ & $7.9115\pm0.5382$& $\textbf{8.6641}\pm\textbf{2.0625}$\\
    FSCS$\uparrow$ & $\textbf{0.6070}\pm\textbf{0.0183}$&$0.5949\pm0.0240$\\		
    \hline\hline
    \end{tabular}
    \label{tab:dis_2}
\end{table}

\subsubsection{Loss Function}
\begin{figure}[ht!]
\centering
\includegraphics[width=2.8in]{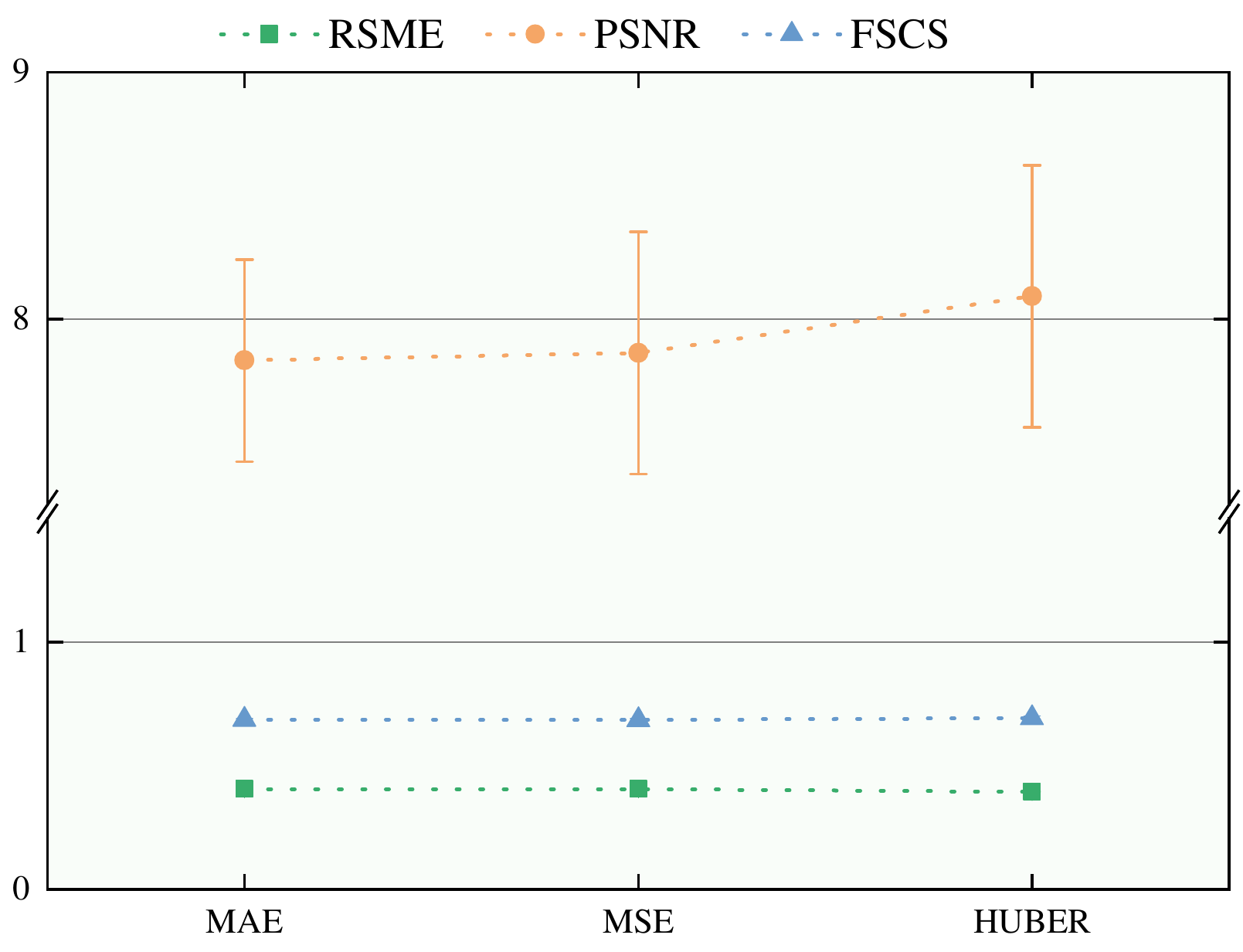}
\caption{Metrics indexes with three loss functions under NC and vary-state condition}\label{loss_dis}
\end{figure}

While typical loss functions often adopt Mean Squared Error (MSE) or Mean Absolute Error (MAE), our proposed method employs the Huber loss, which combines the benefits of both. Nevertheless, we also explored the usage of these two conventional loss functions. Here we adopt same experimental settings under normal condition and SQ vary-state experiment discussed in the previous section. As Fig. \ref{loss_dis} depicts , different loss functions lead to the model converging to a satisfactory result. The utilization of the Huber loss function showcases the most favorable outcomes in terms of PSNR, exhibiting performance improvements compared to MAE and MSE loss.

\subsection{Limitations and Outlook}
The merits of the proposed method can be summarized as follows: the integration of the DMs contributes to a more stable and easily converging training process and incorporation of conditional voltage signals enhances the controllability and detail of generated vibration signals. Moreover, the control condition allows the generation of vibration signals without being limited to specific speeds or scenarios. However, the proposed method also presents certain limitations. For example, it demands a higher GPU memory requirement for encoding control conditions during training compared to other generative methods like GANs. In addition, the reliability of inference for unseen scenarios requires further exploration.

In terms of future development, the proposed method holds promise for data supplementation and feature enhancement, making it applicable to various HSR fault diagnosis tasks such as Remaining Useful Life (RUL) or Anomaly Detection (AD). Moreover, the sample-free inference capability allows for rapid generation of vibration signals using other conditional control signals, not limited to pulse voltage. Exploring new paradigms, such as classifier-free diffusion model generation, could enable comprehensive process monitoring and diagnosis in a cost-effective manner. 
\section{Conclusion}
In order to address the challenges of training instability and dynamical inaccuracy for various HSR fault diagnosis tasks on Internet of Things (IoT) systems, this article proposes an innovative fault vibration signal generation method driven by pulse voltage conditions. This approach involves the joint learning of control features and fault impulse features through generative models, solely sampling from control voltage can generate realistic vibration signals. Leveraging the conditional denoising diffusion model to generate signals from progressively denoising process ensures the stability of training generation. By incorporating features extracted from motor pulse voltage by cross-attention mechanism ensures that the generated vibration signals closely align with real control signals. The performance of our proposed method has been validated across SQ and real HSR bogie bearing dataset, achieving the best RSME, PSNR and FSCS compared with other generative method. Our experimental results attest to the method's robustness and versatility, standing as a powerful tool to overcome diagnostic complexities in scenarios where machines operate under complex and dynamical conditions. Its application can significantly enhance effectiveness and efficiency in downstream HSR fault diagnosis tasks.

% \section*{Acknowledgment}
% The authors would like to thank 

% or


% ============================================
%\appendices
%Appendix one text goes here %\cite{Roberg2010}.

%\section{}

%\section*{Acknowledgment}

%The authors would like to thank D. Root for the loan of the SWAP. The SWAP that can ONLY be usefull in Boulder...

\ifCLASSOPTIONcaptionsoff
  \newpage
\fi



%\begin{thebibliography}{1}


\bibliographystyle{IEEEtran}
\bibliography{IEEEabrv,Bibliography}

\vfill


\end{document}